\documentclass[aps,prl,twocolumn,showpacs]{revtex4}

\usepackage{amssymb,amsmath,graphicx,graphics}

\newcommand{\de}{\partial}
\newcommand{\sech}[1]{\textrm{sech}\left(  #1\right)}
\newcommand{\eq}[2]{\begin{equation} \label{#1} #2 \end{equation}}

\newcommand{\etal}{{\em et al.}}

\begin{document}

\title{Extended temporal Lugiato-Lefever equation and the effect of conjugate fields in optical resonator frequency combs}
\author{Cristian Redondo Lour\'es}
\author{Daniele Faccio}
\author{Fabio Biancalana}
\affiliation{School of Engineering and Physical Sciences, Heriot-Watt University, EH14 4AS Edinburgh, UK}

\begin{abstract}
Starting from the infinite-dimensional Ikeda map, we derive an extended temporal Lugiato-Lefever equation that may account for the effects of the conjugate electromagnetic fields (also called `negative frequency fields'). In the presence of nonlinearity in a ring cavity, these fields lead to new forms of modulational instability and resonant radiations. Numerical simulations based on  the new extended Lugiato-Lefever model show that the negative-frequency resonant radiations emitted by ultrashort cavity solitons can impact Kerr frequency comb formation in externally pumped temporal optical cavities of small size. Our theory is very general, is not based on the slowly-varying envelope approximation, and the predictions are relevant to all kinds of resonators, such as fiber loops, microrings and microtoroids. \end{abstract}

\pacs{42.65.-k, 42.65.Tg, 42.60.Da, 42.65.Sf}
\maketitle


\paragraph{Introduction ---} Kerr frequency combs (KFCs), i.e. light sources with a large number of highly resolved and nearly equidistant spectral lines, are attracting a considerable interest in recent years, due to their important applications in metrology, optical clocks, precision spectroscopy, precision time and distance measurements, attosecond pulse generation and complex nonlinear dynamics, to name just a few \cite{kippenberg1,kippenberg2,kippenberg3}. Physical systems that are able to generate KFCs are microring resonators, microtoroids, crystalline resonators, microspheres, photonic-crystal cavities and optical fiber loops \cite{kippenberg1}. Microring resonators are particularly important in this respect, since they are small-size, low-loss, CMOS-compatible and power efficient devices that can be made of different nonlinear materials and are therefore ideal for on-chip KFC generation.

The main ingredients for an efficient KFC formation have been identified to be four-wave mixing (FWM) and temporal cavity soliton (CS) generation \cite{kippenberg2}. There is currently an intense research activity aiming to maximise the spectral extent of the comb and its coherence, and to understand the experimentally obtained spectra from first principles. Due to the extremely complex dynamical behaviour and stability properties of the propagating CSs and patterns in the resonators, an intense theoretical activity on the mathematical properties of the traditionally used averaged propagation equation, called the temporal Lugiato-Lefever equation (LLE), has developed over the past years, with a frequent display of new and surprising results \cite{kippenberg3,parrarivas1,parrarivas2,parrarivas3,roguewaves}.

One of the major dynamical effects in the propagation of ultrashort pulses is the radiation emitted by solitons due to higher-order dispersion effects, also called resonant radiation (RR) \cite{akhmediev,fission,biancalana,skryabinscience}. This radiation, which appears when pumping near the zero-dispersion point of the structure, is very visible in experiments performed with optical fibers and it also plays a central role in the dynamics of CSs \cite{parrarivas3,skryabin1,erkintalo1,erkintalo3}. Indeed, it has been shown experimentally that CSs emit RR in microring resonators and fiber loops \cite{erkintalo3,gaeta1,gaeta2}.

In addition to the traditional RR emission, other emissions are possible when considering the influence of conjugate fields and the nonlinear interaction between positive and `negative frequencies' in the nonlinear polarization \cite{rubino}. Recently, experiments performed in optical fibers and bulk crystals revealed the presence of `negative frequency resonant radiation' (NRR), which has also been completely explained theoretically in Ref. \cite{biancalananegative}. The theoretical approach used was one based on the analytic signal, which is able to take into account the dynamics of both positive and negative frequencies, thus avoiding the use of the slowly-varying envelope approximation which breaks down for very short pulses \cite{biancalananegative,amir1,amir2}. An intriguing question is whether nonlinear optical cavities may somehow enhance, and therefore be affected by NRR-like effects.

In this paper we extend the temporal LLE in order to take into account the effects of negative frequencies and conjugate fields on the propagation of CSs in optical resonators ({\em extended LLE}, eLLE for brevity). This extension is necessary in order to be able to study analytically and numerically the existence of new resonant radiations emitted by the CSs during the circulation in the resonator. We show that due to the forced and dissipative nature of the eLLE, the new resonant radiations, that are typically remarkably feeble in conventional optical fibers and bulk materials, can become quite strong and can be more efficiently generated. This surprising result has the potential to impact considerably the formation of KFCs due to CSs and also the intrinsic stability of the homogeneous steady state CW solutions of the cavity.

\paragraph{Extended temporal Lugiato-Lefever equation ---} 
The starting point of any discussion on optical resonators and cavities is the infinite-dimensional Ikeda map \cite{haelterman1,haelterman2}
\begin{eqnarray}
A_{n+1}(0,t)=TA_{\rm in}(t)+Re^{-i\phi_{0}}A_{n}(L,t),\label{eq1}\\
i\de_{z}A_{n}+i\frac{\alpha_{\rm i}}{2}A_{n}+\hat{D}(i\de_{t})A_{n}+\hat{S}(i\de_{t})p_{\rm nl}[A_{n}]=0.\label{eq2}
\end{eqnarray}
Here, $A_{\rm in}$ is the envelope of the (impulsed or cw) pump field, $T$ is the transmission coefficient at the coupling point $z=0$, $R$ is the reflection coefficient at the same point ($R^{2}+T^{2}=1$), $\phi_{0}=\delta_{0}-2\pi m$ is the phase accumulated over a round-trip, $\delta_{0}$ is the cavity detuning, $L$ is the length of the cavity, $A_{n}(z,t)$ is the envelope of the intracavity field circulating at the $n$-th step, $z$ is the spatial coordinate along the cavity, $t$ is the `fast' time variable in the reference frame moving at the group velocity $1/\beta_{1}$, $\hat{D}(i\de_{t})\equiv\sum_{j\geq 2}\beta_{j}(i\de_{t})^{j}/j!$ is the dispersion operator, $\beta_{j}\equiv[\de_{\omega}^{j}\beta(\omega)]_{\omega=\omega_{0}}$ is the $j$-th dispersion coefficient calculated at the pump frequency $\omega_{0}$, $\hat{S}(i\de_{t})\equiv(1+i\omega_{0}^{-1}\de_{t})$ is the shock operator describing the first order correction due to the frequency dependent nonlinearity. For simplicity, we do not consider here the influence of the Raman effect, as this can also be easily included in our model.

The Ikeda map (\ref{eq1}-\ref{eq2}) is usually based on the slowly-varying envelope approximation (SVEA), under which the spectral extent of the pulse must be much smaller than its central frequency. In Eq. (\ref{eq2}), $p_{\rm nl}$ is the nonlinear polarization of the intracavity field, which in presence of the Kerr effect only can be written as $p_{\rm nl}[A_{n}]=\gamma|A_{n}|^{2}A_{n}$, where $\gamma$ is the nonlinear coefficient of the material. However, as was explained in Ref. \cite{biancalananegative}, this simple form of the polarization does not describe sufficiently well the dynamics of ultrashort pulses. In order to include in a physically consistent way the higher-order effects such as third-harmonic generation (THG) and the contribution of conjugate terms, one must use the so-called {\em analytic signal} (AS) for all the fields involved in Eqs. (\ref{eq1}-\ref{eq2}). We refer the reader to Refs. \cite{biancalananegative,amir1,amir2} for a full discussion, but we review here the salient features of ASs for completeness.

\begin{figure}
\includegraphics[width=5cm]{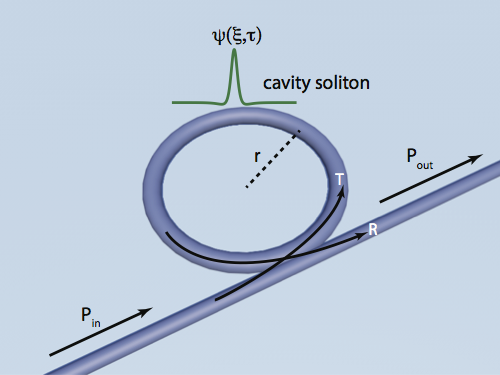}
\caption{(Color online) Sketch of the generic microcavity and its parameters. $r$ is the ring radius, $P_{\rm in}$ is the cw input power, $P_{\rm out}$ is the output field power, $\psi(\xi,\tau)$ is the intracavity field at position $\xi$ in the ring, $R$ and $T$ are the reflection and transmission coefficients, respectively.}
\label{fig1}
\end{figure}

The full real electric field propagating in an optical system is denoted by $E(z,t)$, where $z$ is the propagation direction and $t$ is the time variable. Its Fourier transform is denoted by $E_{\omega}(z)\equiv\mathcal{F}[E(z,t)]=\int_{-\infty}^{+\infty}E(z,t)e^{i\omega t}dt$. The analytic signal of the electric field, i.e. the positive frequency part of the field, is defined as $\mathcal{E}(z,t)\equiv \pi^{-1}\int_{0}^{\infty}E_{\omega}(z)e^{-i\omega t}d\omega$, and is a complex function. The Fourier transform of the electric field can be written as the sum $E_{\omega}=[\mathcal{E}_{\omega}+(\mathcal{E}_{-\omega})^{*}]/2$ since only the positive (or negative) frequency part of the spectrum carries information, while for the same reason the electric field itself is real and is given by $E(z,t)=[\mathcal{E}(z,t)+\mathcal{E}^{*}(z,t)]/2$.
The `envelope' we introduce here is defined by $A(z,t)\equiv\mathcal{E}(z,t)e^{-i\beta_{0}z+i\omega_{0}t}$ (where $\beta_{0}$ is the propagation constant at the pump frequency), i.e. the frequency components of the AS are `shifted' by an amount $-\omega_{0}$. By doing this, we deal with frequency detunings $\Delta\omega$ from $\omega_{0}$, and not with absolute frequencies. However, {\em there is a key difference} between the conventional definition of envelope and the envelope of the AS: the former is adequate only if the spectral extension of the pulse evolution is much smaller than the pulse central frequency, $|\Delta\omega|\equiv|\omega-\omega_{0}|\ll\omega_{0}$, i.e. only under SVEA conditions, while the envelope of the AS $A(z,t)$ considered here does not suffer from this limitation. By clearly dividing the envelope associated to the positive frequency components from that associated to the negative ones, we will be able to write the extended temporal LLE that correctly describes the dynamics of intracavity pulses {\em of arbitrary duration and spectral extension}, taking into account the peculiar and non-trivial interaction between positive and negative frequencies that arises in the complete nonlinear polarization.
In this paper, all fields, linear and nonlinear, are intended to be envelopes of ASs.

Within the above framework, the AS of the nonlinear polarization is written as \cite{biancalananegative}:
\eq{pnl1}{p_{\rm nl}[A_{n}]=\gamma\left[ |A_{n}|^{2}A_{n} + |A_{n}|^{2}A_{n}^{*}e^{2i\phi(z,t)} + \frac{1}{3}A_{n}^{3}e^{-2i\phi(z,t)} \right]_{+},} where $\phi(z,t)\equiv\omega_{0}t+\Delta kz$, and $\Delta k\equiv(\beta_{1}\omega_{0}-\beta_{0})$ is a factor, crucial for the efficient phase-matching of the resonant negative-frequency terms, that measures the difference between the phase and the group velocity in the medium.
The subscript $+$ in Eq. (\ref{pnl1}) signifies the filtering of the negative frequency components out of the polarization. This ensures that during its evolution $A_{n}(z,t)$ only contains positive frequencies, and it is thus consistent with its own definition. The first term in Eq. (\ref{pnl1}) is the usual Kerr term, the second is the so-called 'negative Kerr' (NK) term, while the third gives THG. Note that NK and THG {\em must} appear together in order to preserve the Hamiltonian nature of the FWM interaction, something that was overlooked prior to our theoretical work on the subject \cite{biancalananegative}.

The accurate numerical integration of Eqs. (\ref{eq1}-\ref{eq2}) is very time consuming, and the equations are not suitable for a deep analytical understanding. This is the reason why mean-field models based on averaging over many roundtrips (see Ref. \cite{lugiatolefever}) are a common tool for dramatically reducing the complexity of the map into a single equation. To this aim, we now plug Eq. (\ref{pnl1}) into Eq. (\ref{eq2}), and we follow the averaging procedure described in \cite{haelterman2}, using the `cavity soliton units' $\xi\equiv z/L_{\rm D2}$, $\tau\equiv t/t_{0}$, $\Omega\equiv\omega_{0}t_{0}$, $\kappa\equiv\Delta kL_{\rm D2}$, $\psi\equiv A/\sqrt{P_{0}}$, $P_{0}\equiv(\gamma L_{\rm D2})^{-1}$ (so that $L_{\rm NL}\equiv[\gamma P_{0}]^{-1}=L_{\rm D2}$), $\psi_{\rm in}\equiv A_{\rm in}/\sqrt{P_{0}}$, $\phi(\xi,\tau)\equiv\Omega\tau+\kappa\xi$. A natural value for the time scale $t_{0}$ is the typical single-CS duration $t_{0}=[|\beta_{2}|L/(2\delta_{0})]^{1/2}$. In the averaging procedure, the cavity length $L$ must be much smaller than the dispersive and nonlinear lengths, $L_{\rm D2}$ and $L_{\rm NL}$ respectively. This ensures that the intracavity pulse does not change much during a single roundtrip. This slow variation must also be satisfied by the NK and THG terms in Eq. (\ref{pnl1}). This means that the phase $\phi(z,t)$ must rotate very rapidly in a single roundtrip, so that the average effect is mediated {\em almost} to zero, resulting in the condition $\kappa\gg 1$, meaning that the group velocity and the phase velocity must be sufficiently different, which is true in dispersive media such as dielectric waveguides. 

In this way, we arrive at the following {\em extended Lugiato-Lefever equation} (eLLE): \begin{widetext}
\eq{governing1}{i\de_{\xi}\psi+\hat{D}(i\de_{\tau})\psi+i(\Gamma+i\delta)\psi+(1+i\Omega^{-1}\de_{\tau})\left[|\psi|^{2}\psi + |\psi|^{2}\psi^{*}e^{2i\phi(\xi,\tau)} + \frac{1}{3}\psi^{3}e^{-2i\phi(\xi,\tau)}\right]_{+}-i\mu\psi_{\rm in}=0.}
\end{widetext}
$\Gamma\equiv[(\alpha_{\rm i}L+T^{2})/2]L_{\rm D2}/L$, $\delta\equiv\delta_{0}L_{\rm D2}/L$, $\mu\equiv TL_{\rm D2}/L$, $\hat{D}(i\de_{\tau})\equiv\sum_{n\geq2}b_{n}(i\de_{\tau})^{n}$, where $b_{n}\equiv\beta_{n}/(n!t_{0}^{n-2}|\beta_{2}|)$ are the dimensionless dispersion coefficients, and $\phi(\xi,\tau)\equiv\Omega\tau+\kappa\xi$.

Equation (\ref{governing1}) is the central result of this paper. It can be applied to any kind of optical resonator, and can be easily extended to include the Raman effect or any other perturbation of the NLSE.

\paragraph{Phase matching conditions for the new radiations ---} We now derive the phase-matching conditions for the most important resonant radiations emitted by the ultrashort CSs propagating in the resonator. In Eq. (\ref{governing1}), we substitute the {\em Ansatz} $\psi(\tau,t)=\psi_{0}+g(\xi,\tau)$, and linearize with respect to the small radiation field $g$. Note that due to the loss term proportional to $\Gamma$ in Eq. (\ref{governing1}), far from the central peak of the CS the resonant radiations decay asymptotically towards the complex cw background $\psi_{0}$. Thus the intensity of the CS peak power does not play any important role, and the problem becomes very similar to a modulational instability (MI) phase-matching. However, the usual MI analysis does not work here, since the presence of the exponential terms in Eq. (\ref{governing1}) cannot give a stationary state, and the standard techniques used in Refs. \cite{haelterman1,haelterman2,skryabin1} would not be appropriate. We therefore use the perturbation method employed in Ref. \cite{biancalananegative} in the fiber-optics/bulk context, obtaining \begin{eqnarray}
D(\Delta)-v\Delta=D_{0}, \label{pm1} \\
D(\Delta)-v\Delta=D_{0}\pm2\kappa, \label{pm2}
\end{eqnarray}
where $\Delta$ is the dimensionless frequency detuning from the pump, $D(\Delta)\equiv\sum_{n\geq2}b_{n}\Delta^{n}$ is the dispersion of the linear waves, $D_{0}\equiv\delta-i\Gamma-2|\psi_{0}|^{2}$ is the complex nonlinear wavenumber, and $v$ is the velocity parameter of the resulting moving CS, which starts to drift as a consequence of the higher-order dispersion \cite{skryabin1}. Equation (\ref{pm1}) is connected to the emission of the conventional RR, routinely observed in fibers and cavities \cite{biancalana,skryabinscience,erkintalo3}. Equations (\ref{pm2}) phase-match the {\em negative frequency resonant radiation} [NRR, `+' sign in Eq. (\ref{pm2})], which is due to the NK term in Eq. (\ref{pnl1}), and the {\em third-harmonic resonant radiation} [THRR, `--' sign in Eq. (\ref{pm2})], which is due to the THG term in Eq. (\ref{pnl1}). 


Note that in all physically relevant situations $\kappa\gg|D_{0}|,|v\Delta|$. The roots of Eqs. (\ref{pm1}-\ref{pm2}) have real and imaginary parts, indicating the frequency detuning of the emissions and their decay rate towards the soliton background, respectively. Equation (\ref{pm1}) is valid only in the presence of a propagating CSs, while Eqs. (\ref{pm2}) holds also for a pure cw intracavity field. This latter phase-matching equation implies that the `old' stable branch of the stationary solutions of Eq. (\ref{governing1}), i.e. the smallest root of the equation $i(\Gamma+i\delta)\psi+|\psi|^{2}\psi-i\mu\psi_{\rm in}=0$ (see also Refs. \cite{barashenkov,erkintalo2}), is unstable in presence of NK and THG, and sidebands will appear. This effect, which is not present in the traditional temporal LLE, is also a new feature of our eLLE model, as we shall see shortly.

\paragraph{Numerical simulations ---} In order to illustrate the spectral dynamics and the emission of the resonant radiations from ultrashort CSs as seen in the previous section, we use a highly-nonlinear, small, low-loss resonator (for example a microtoroid or a microring, of the kind described in Refs. \cite{kippenberg1,kippenberg2}, see also Fig. \ref{fig1}). We take a radius $r=30$ $\mu$m, cavity length $L=1.88\times 10^{-4}$ m, pump wavelength $\lambda_{0}=1.55$ $\mu$m, $\beta_{2}=-90$ psec$^{2}$/km, $\beta_{3}=-1.11$ psec$^{3}$/km, nonlinear coefficient $\gamma=1$ W$^{-1}$m$^{-1}$, group index at the pump $n_{\rm g}=1.5$, cw pump power $P_{\rm in}=265$ mW, transmission coefficient $T=0.07$, detuning from the cavity resonance $\delta_{0}=0.0115\ll\pi$, free spectral range FSR=$1060$ GHz, roundtrip time $t_{\rm R}=1/$FSR$=150$ fsec, photon lifetime $t_{\rm ph}\simeq 0.19$ nsec, finesse $\mathcal{F}\simeq 628$ and loaded $Q$-factor $Q\sim10^{5}$. The typical temporal width of the CSs formed in the cavity is $t_{0}=[|\beta_{2}|L/(2\delta_{0})]^{1/2}\simeq 27$ fsec, and we take this value for the scaling of Eq. (\ref{governing1}). The typical peak power of the CS is also given by the soliton power scale $P_{0}=2\delta_{0}(L_{\rm D2}/L)=[\gamma L_{D2}]^{-1}\simeq122$ W. The second order dispersion length is $L_{\rm D2}=8.2$ mm, which gives a ratio $L_{\rm D2}/L\simeq 43.5\gg 1$, making the averaging procedure meaningful.
We take a value $\Delta k\sim0.67\times10^{4}$ m$^{-1}$, an order of magnitude that is common for solid media.
It is very favourable to use small-size microcavities in order to minimize the value of $\phi$ accumulated over a roundtrip, proportional to $\Delta kL_{\rm D2}\cdot(L/L_{\rm D2})=\Delta kL$. This gives the opportunity to observe the new nonlinear effects in realistic systems.

With these parameters, the dimensionless coefficients in Eq. (\ref{governing1}) become $\Gamma=0.217$, $\delta=0.5$, $\mu=3$, $\psi_{\rm in}=0.047$, $b_{2}=\pm0.5$, $b_{3}=-0.0758$, $\Omega=33$ and $\kappa\simeq54.5$. Due to the fast oscillations in the exponentials in Eq. (\ref{governing1}), which can generate high-frequency radiations, we make sure to use a large number of sampling points in the split-step Fourier solver ($N=2^{17}$), and a small spatial step ($\Delta\xi=10^{-6}$).

\begin{figure}
\includegraphics[width=8.6cm]{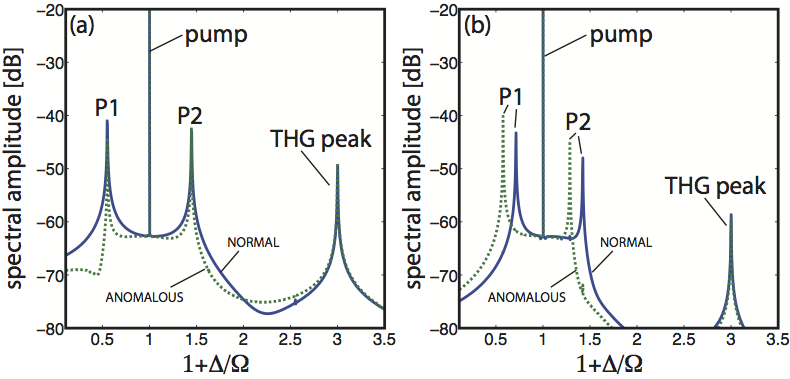}
\caption{(Color online) (a) Output spectrum of the propagation of the HSS cw solution for normal ($b_{2}=+0.5$, blue solid line) and anomalous ($b_{2}=-0.5$, green dashed line) dispersions, in the case $b_{3}=0$. Propagation distance is $\xi=27$. (b) Same as (a) but with $b_{3}=-0.0758$. All other relevant parameters are given in the text.}
\label{fig2}
\end{figure}

Figure \ref{fig2}(a) shows the output spectrum of the propagation ($\xi=27$, equivalent to $1200$ cavity roundtrips) of the lower-branch homogeneous steady state (HSS) cw solution (see Refs. \cite{barashenkov,erkintalo2}) in both anomalous and normal dispersions respectively ($b_{2}=\pm0.5$), by using the eLLE, Eq. (\ref{governing1}), when $b_{3}=0$. As discussed in the previous section, for these parameters the HSS solution is stable in absence of NK and THG terms. Conversely, accounting the complex conjugate fields leads to far-detuned sidebands (indicated by P1 and P2 in the figure) according to the phase-matching conditions of Eq. (\ref{pm2}). Note that this new form of instability appears irrespective of the sign of $b_{2}$, due to the $\pm$ in Eq. (\ref{pm2}). Moreover, the sidebands would be perfectly symmetric with respect to the pump frequency for vanishing higher-order dispersion terms, but become asymmetric when the contribution of $b_{n\geq 3}$ is important, see Fig. \ref{fig2}(b). In this case, the two peaks are always slightly imbalanced in amplitude, due to the presence of THG and pump and loss in the system. This proves, for the first time to our knowledge, the existence of novel kinds of modulational instability that are due to the contribution of negative frequencies in the nonlinear polarization [Eq. (\ref{pnl1})]. This MI can be of the order of 1\% of the input pump or larger and should therefore be readily visible in experiments.

\begin{figure}
\includegraphics[width=8cm]{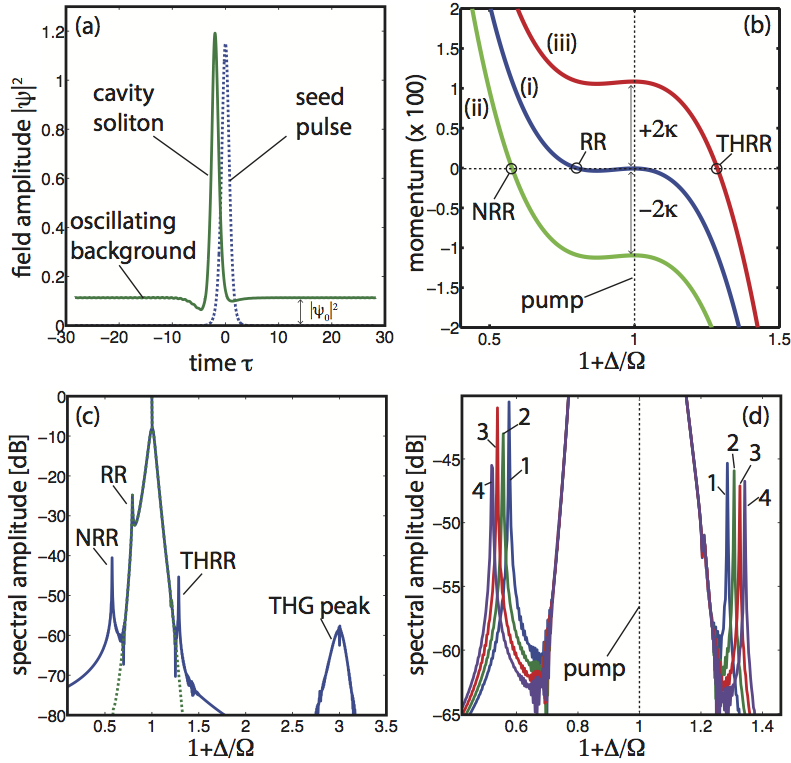}
\caption{(Color online) (a) Formation of a stable moving CS in the cavity, after a propagation of $\xi=27$. (b) Real part of phase-matching curves for the three equations (\ref{pm1}-\ref{pm2}): (i) for Eq. (\ref{pm1}), (ii) for Eq. (\ref{pm2}) with $+$ sign, and (iii) for Eq. (\ref{pm2}) with $-$ sign. Circles are the predicted positions of all the generated radiations (RR, NRR, THRR). (c) Intracavity field spectrum at $\xi=27$, when considering only the Kerr term (green dashed line), and when including the NK and the THG terms in Eq. (\ref{pnl1}) (blue solid line). The resonant radiations generated by the NK and the THG terms are indicated with NRR and THRR, respectively. (d) Zoom of the NRR and THRR peaks when $\kappa=54.5, 65, 76, 87$ (indicated by 1,2,3,4 respectively). The parameters of the simulation can be found in the text, and $b_{2}=-0.5$, $b_{3}=-0.0758$, $v=0.084$.}
\label{fig3}
\end{figure}

Figure \ref{fig3}(a) shows the formation of a stable moving CS in the cavity, in presence of third-order dispersion, $b_{3}\neq0$ and in anomalous dispersion $b_{2}=-0.5$, when using the eLLE Eq. (\ref{governing1}). The input seed is taken to be $\psi(\xi=0,\tau)=\sqrt{2\delta}\sech{\sqrt{2\delta}\tau}$, i.e. the autosolution of the pulsed cavity \cite{erkintalo2}, which is a good approximation (albeit with vanishing background) of the final CS. 
Figure \ref{fig3}(b) shows the real part of the phase-matching curves of Eqs. (\ref{pm1}-\ref{pm2}), and the prediction of the frequency positions of all the resonant radiations.
Green dashed line in Fig. \ref{fig3}(c) shows the intracavity field spectrum, in the case when only the Kerr effect is present in the nonlinear polarization of Eq. (\ref{pnl1}) (i.e. the case of the conventional temporal LLE). In this case, only the usual RR peak, which satisfies Eq. (\ref{pm1}), is emitted near the pump (with a very small contribution of its symmetric counterpart, not discussed here, see \cite{skryabin1}). Blue solid line in Fig. \ref{fig3}(c) shows the intracavity field spectrum when taking into account all the terms in the full polarization, which contains also the NK and the THG terms, see Eq. (\ref{pnl1}). One can notice the appearance of relatively strong peaks, which are the NRR and the THRR peaks described in the previous section, and that satisfy the phase-matching conditions Eqs. (\ref{pm2}). Complete agreement is observed between the predicted radiation positions [Fig. \ref{fig3}(b)] and the simulated ones. Also note the appearance of the THG peak, located at $1+\Delta/\Omega=3$. 
Figure \ref{fig3}(d) shows the position of the NRR and THRR peaks for different values of $\kappa$. The amplitudes of the radiations are uneven since the CS background oscillates during propagation.

\paragraph{Discussion and conclusions ---} Our results in this work show that frequency comb formation in Kerr media is affected by the resonant radiations and MI resulting from the nonlinear interactions between positive and negative frequencies. These new types of radiations play a conceptually important but somewhat minor role in optical fibers and bulk materials. However, in optical resonators such as fiber loops and microrings, the field is forced to circulate many times in the cavity, and CSs possess a cw background that may dramatically enhance the emission of the new radiations. As a consequence, the output KFC spectra are affected by the radiation peaks emitted by CSs, and the `negative frequency' radiations play an important role. We have provided the derivation of a universal model (the eLLE), {\em not based on SVEA}, that is able to take into account the full dynamics of the ultrashort intracavity pulses, and is amenable to analytical investigations.

Different alternative (but equivalent) approaches exist in the description of KFC formation \cite{matsko1,chembo1}. In one scheme, the dynamics of the annihilation operators of the cavity modes is described via Heisenberg equations. In the same spirit of our formulation, the inclusion of the negative-frequency effects could be accounted for by using the complete (and correct) interaction Hamiltonian $V=-(g\hbar)[e^{\dagger}e^{\dagger}ee/2+(e^{\dagger}eee+e^{\dagger}e^{\dagger}e^{\dagger}e)/3]$, in the language of Ref. \cite{matsko1}. The first term is the usual FWM scattering, while the second and third terms, which must come together since they are individually non-Hermitian, account for THG and NK effects, respectively.
It would be interesting to see how the different approaches in the literature may converge towards a unified description of the phenomena described here.

F.B. would like to thank S. Trillo, N. Joly, M. Erkintalo, M. Clerici and L. Caspani for useful discussions.

\end{document}